# Temperature Invariant Metasurfaces

Shany Zrihan Cohen[1,2], Danveer Singh[1,2], Sukanta Nandi[1,2] and Tomer Lewi[1,2]

[1]Faculty of Engineering, Bar-Ilan University, Ramat-Gan 5290002, Israel
[2]Institute for Nanotechnology and Advanced Materials, Bar-Ilan University Ramat-Gan 5290002, Israel
*Corresponding authors:  tomer.lewi@biu.ac.il

**Abstract**

Thermal effects are well known to influence the electronic and optical properties of materials through several physical mechanisms and are the basis for various optoelectronic devices. The thermo-optic (TO) effect–the refractive index variation with temperature (dn/dT), is one of the common mechanisms used for tunable optical devices, including integrated optical components, metasurfaces and nano-antennas. However, when a static and fixed operation is required, i.e., temperature invariant performance - this effect becomes a drawback and may lead to undesirable behavior through drifting of the resonance frequency, amplitude, or phase, as the operating temperature varies over time. In this work, we present a systematic approach to mitigate thermally induced optical fluctuations in nanophotonic devices. By using hybrid subwavelength resonators composed from two materials with opposite TO dispersions (dn/dT<0 and dn/dT>0), we are able to compensate for TO shifts and engineer meta-atoms and metasurfaces with zero effective TO coefficient (dn/dT≈0). We demonstrate temperature invariant resonant frequency, amplitude, and phase response in meta-atoms and metasurfaces operating across a wide temperature range and broad spectral band. Our results highlight a path towards temperature invariant nanophotonics, which can provide constant and stable optical response across a wide range of temperatures and be applied to a plethora of optoelectronic devices. Controlling the sign and magnitude of TO dispersion extends the capabilities of light manipulation and adds another layer to the toolbox of optical engineering in nanophotonic systems.

**Introduction**

The thermo-optic (TO) coefficient is a macroscopic property defining the refractive index variation with temperature (dn/dT) and can be assigned to materials, composites, and devices [1]. Many optoelectronic applications utilize this effect as a tuning [2–9], modulating [10–14], or sensing mechanism [15–18], where small changes in temperature lead to resonance wavelength shifts, variations in phase propagation, or other optical properties. Indeed, a plethora of active TO based devices have been successfully implemented in fiber-based devices [19,20], integrated optics [16,21] and metasurfaces [2,3,5,6,22–24]. Although the benefits of the TO effect are obvious, when a stable and fixed optical response is required, it may become a drawback as the optical properties of the device fluctuate with temperature, leading to undesirable behavior. In high-Q resonant structures such as ring resonators with typical linewidth of ~ 10 picometers, temperature changes as small as 0.1K will shift the operating resonance wavelength by 40 picometers [21,25] (i.e., 4 times the resonance linewidth), which would be detrimental to the device operation. Similar effects can also occur in low-Q resonant structures which are subjected to large temperature gradients (10s or 100s of kelvin degrees), where the optical properties may change dramatically. Extreme temperature gradients inherently exist in space applications [26], however, large and moderate temperature variations may also occur in laser systems [27], thermophotovoltaics, photodetection and sensing [17,28]. Temperature variations may be localized - due to, e.g., local heating from intense light sources [4] or electronic circuits, or non-local, due to changes in the ambient temperature, as in the case for space applications or sensors.

Consider for example the temperature dependent mid-infrared (MIR) spectra of a silicon disk metasurface shown in Figure 1a, plotted for the temperature range 143-643K. The spectra exhibit strong temperature dependent resonance red shifts ($\Delta\lambda \approx 200$nm, 0.4 nm/K) which ultimately result in significant changes in all the optical properties (i.e., amplitude, phase). Such strong temperature dependency is a result of the large TO coefficient of silicon ($dn_{Si}/dT = 2.5 \cdot 10^{-4}/°K$ [1,5]) and cannot be ignored when the device is subjected to large temperature variations (as shown in Figure 1a). Figure 1b, presents FDTD calculated reflection of a high-Q metasurface designed around Fano-resonances with broken symmetry unit-cell [29] for the NIR spectral range. In this high-Q metasurface (Q~285), even moderate temperature variations of ~40K will dramatically change the reflection of the device (by ~90%). Thus, the effect of thermo-optically induced shifts becomes increasingly significant for high-Q structures, and particularly in ultra-high-Q resonant structures such as microring, microsphere, and microtoroid resonators. Even small temperature variations, in the range of fractions of a kelvin degree, can have a detrimental impact on the performance of these devices.

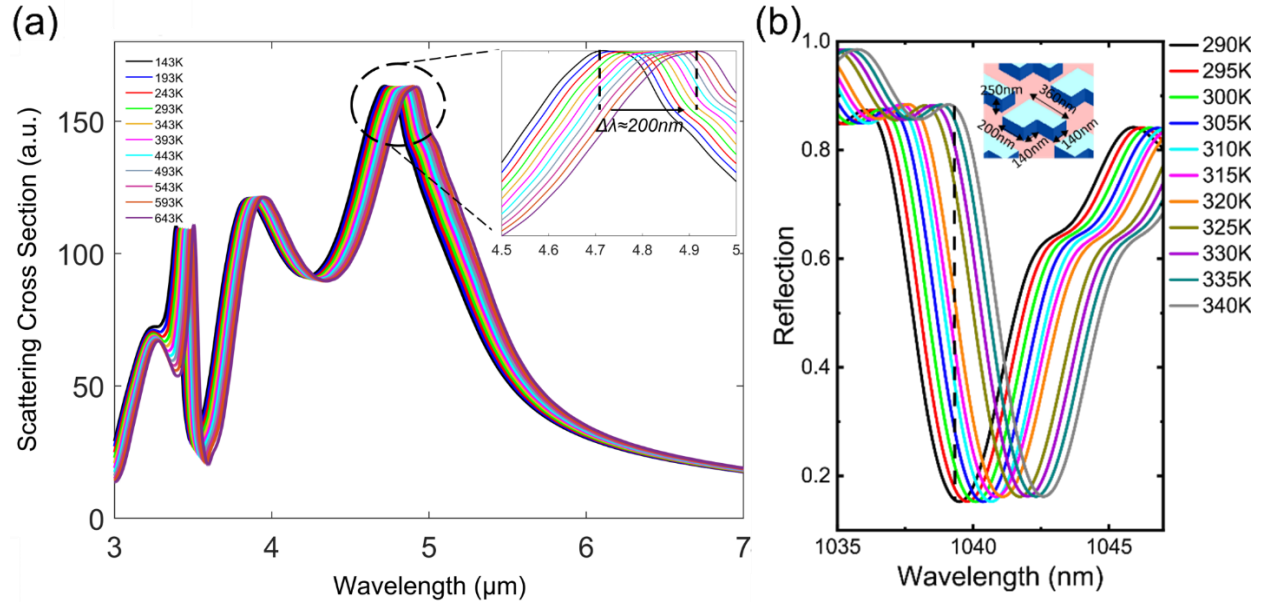

*Figure 1*: Thermo-optic induced shifts in low and high-Q metasurfaces. (a) Low-Q Metasurface comprised of Silicon disks exhibiting large red shift (Δλ≈200nm) with temperature. The disk dimensions are d (diameter)= 1.3µm, h (height)= 1.2 µm and the periodicity is Λ = 2.2µm. The substrate is BaF$_2$. (b) High-Q Fano resonant metasurface with unit cell dimensions depicted in the inset. The meta-atoms are silicon atop of a SiO$_2$ substrate. Moderate temperature variations of ~40K dramatically change the reflection of the device

Temperature and thermodynamic effects in metasurfaces have been previously studied in space applications [30], laser science [31], thermal emission [30,32,33] and photodetection, where eliminating temperature effects in the performance of nanophotonics devices has been only studied in thin film geometry [34,35].

In this work, we present a systematic approach for temperature invariant metasurfaces that retain their optical properties over a wide temperature range. By using hybrid resonators composed from two materials with opposite TO dispersions (dn/dT<0 and dn/dT>0), we compensate for TO shifts and engineer meta-atoms and metasurfaces with zero effective TO coefficient (dn/dT≈0). These TO engineered metasurfaces maintain static and fixed optical response with respect to frequency, amplitude, and phase even for temperature gradients as large as 500K degrees. We demonstrate this approach for a variety of typical nanophotonic components including Bragg mirrors, single spherical, cubic and disk resonators and finally for large metasurface arrays. Our findings demonstrate that by controlling the sign and magnitude of TO dispersion, it is possible to cancel or mitigate thermally induced shifts in optical systems, leading to increased stability, efficiency, and performance. This approach expands our capability to manipulate light and may broaden the potential applications in nanophotonic systems.

**RESULTS**

We divide our results to 1D Bragg mirrors, single Mie resonators of various geometries, and full metasurfaces. To compensate for TO induced optical shifts we use hybrid unit-cell meta-atoms with positive and negative TO coefficients. For our positive TO material (dn/dT>0), we selected Silicon – a high index material which is transparent across the infrared range. To compensate for the positive TO coefficient of silicon, the structure must include materials with negative TO coefficients. While most high index dielectrics and semiconductors possess positive TO coefficients, the lead chalcogenide family (PbTe, PbSe, PbS) has both high refractive indices and an anomalous negative TO effect (see supporting information section 1 for more details). We selected PbTe as the negative TO material, as it has the highest refractive index and largest (in magnitude) TO coefficient among the lead chalcogenides [1,5]. These two materials, Silicon and PbTe, form the basic hybrid resonator unit cells in our structures. In our designs, we also take into consideration the chromatic dispersion (n(λ)) of the refractive indices of the materials.

**1D structures: Bragg Mirrors**

Bragg mirrors are widely used in various applications due to their high reflectivity and low losses [35,36] . Standard Bragg mirrors incorporate stacks of alternating layers of high and low refractive indices and can be designed for a specific central wavelength as well as to function as bandpass or bandstop filters. Here, we implement the same approach, however, the layers alternate not only in the refractive index, but also in the sign of the TO coefficient. PbTe is the high refractive index material ($n_{PbTe}$ =5.85 @5μm) and has a negative thermo-optic coefficient $dn_{PbTe}/dT$ = -13.5 [$10^{-4}$/K] [1,5], whereas Si acts as the lower refractive index layer ($n_{Si}$≈3.42 in the infrared range) and has a positive TO coefficient $dn_{Si}/dT$=2.5 [$10^{-4}$/K] [1].

The infrared spectra of an 11-layer stack are presented in Figure 2. While all resonances across the broad 3-12 µm spectrum exhibit very low temperature dependency, the results were optimized for the central wavelength $\lambda_0$ = 5.36µm. Indeed, at this resonance wavelength, the structure demonstrates extremely high reflectivity (R ≈ 0.99) for all the simulated temperatures, throughout a wide bandpass Δλ ≈ 880nm, as well as temperature invariant performance across a 500K temperature span, with no spectral shift of the central resonance wavelength.

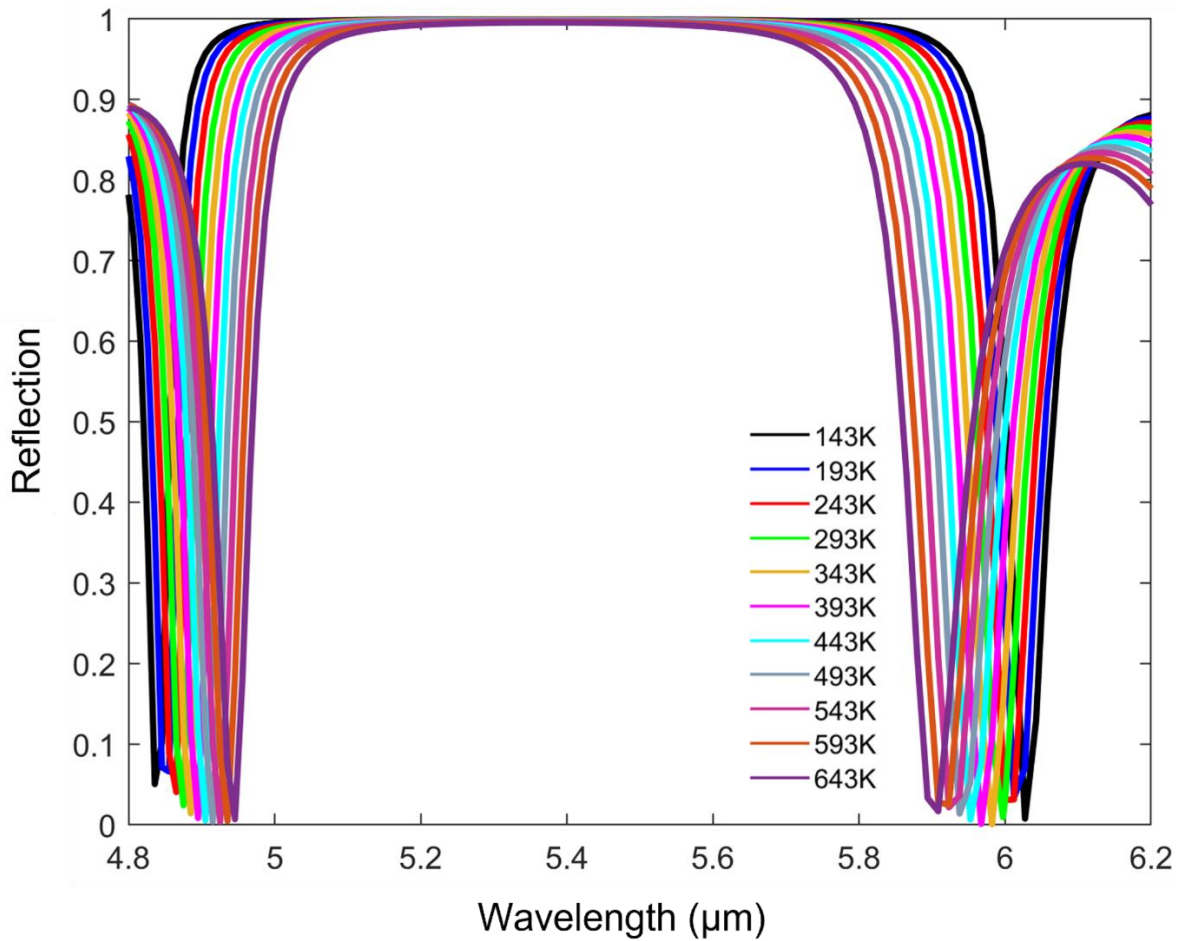

**Figure 2**: Temperature independent Bragg mirror incorporating 11 alternating layers of PbTe – Si with thicknesses $d_{PbTe}$ = 0.19µm, $d_{Si}$ =1.25µm. The central resonance wavelength λ = 5.36µm exhibits high reflectivity as well as temperature invariance behavior

Interestingly, we observe bandpass narrowing of $\Delta\lambda_{max} \approx 125$nm as temperature increases, which is a result of the reduction in refractive index contrast with temperature; namely, $n_{PbTe}$ decreases with temperature, while $n_{Si}$ increases.

**2D structures: Three-layer Hybrid Mie Resonators**

The power in manipulating free space light lies in two-dimensional metastructures. Mie resonator meta-atoms form the basic unit cell building blocks for metasurfaces and metamaterials [37] and are also excellent scatterers as single or ensembles of nano-antennas [38]. We start by designing temperature invariant hybrid single Mie-resonators spanning various geometries. We select three common geometries – sphere, disk, and cubic resonators, and carefully engineer each of the structures to eliminate the thermal dispersion ($dn_{eff}\backslash dT \approx 0$) for a given mode. The structures are symmetric in the x-y plane, and the incident beam is x-polarized propagating along the z-axis. The unit cell in all structures is composed of three Si/PbTe/Si layers. The scattering spectra at all temperatures were obtained using finite difference time domain (FDTD) solver (see supporting information section 3).

1. Spherical multilayer core-shell Hybrid Mie Resonators

The spectra of multilayer core-shell Si/PbTe/Si spherical Mie resonator is presented in Figure 3a (the spectra are vertically shifted along the y-axis for visibility). Optimization of the structure was done through parameters sweep (radii and core-shell thicknesses) and resulted in very small and negligible variations of the resonant frequency, across a wide spectral and temperature ranges. Namely, the spectral position of the first three resonant modes: the magnetic dipole (MD), electric dipole (ED), and magnetic quadrupole (MQ), are fixed. In addition to the spectral response, both the phase and the amplitude remain the same, as presented in Figure 3b. This is exemplified by the overlapping amplitude and phase response across the 6-10 µm spectral range, at 143K and 643K, respectively. To quantify the resonance wavelength shifts, we define a normalized figure of merit (FOM) $S_{max} = \Delta\lambda/\lambda_0$ where $\lambda_0$ is the resonance wavelength at room temperature (RT), and $\Delta\lambda$ is the maximum resonance shift compared to the resonance wavelength at RT. Figure 4 present the extracted temperature dependent resonance wavelength of the MD, ED and MQ modes of Figure 3a.

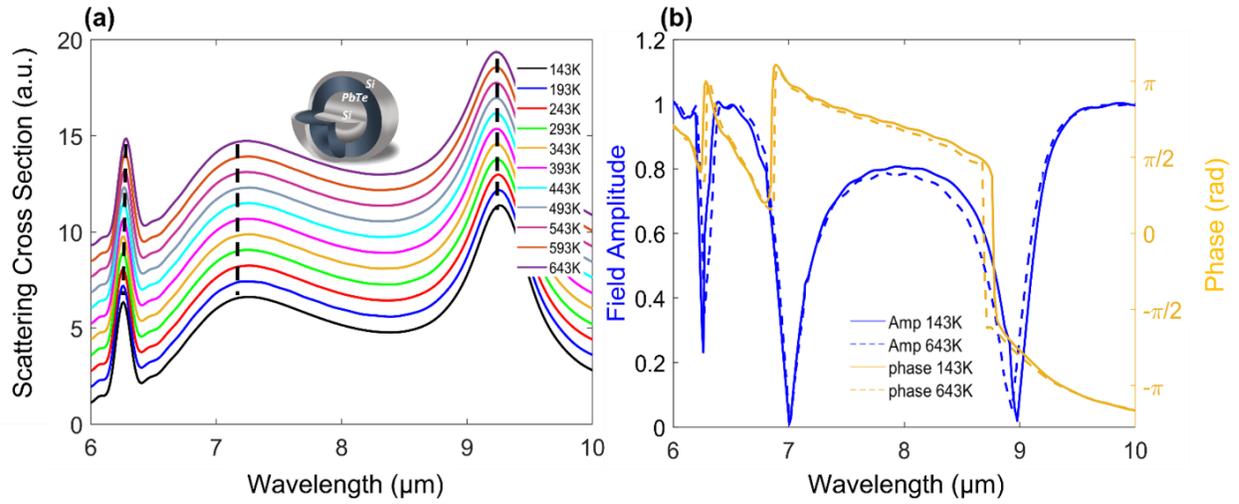

**Figure 3**: (a) Scattering spectra of single multilayer spherical resonator, demonstrating fixed spectral position across a ΔT=500K temperature swing. The hybrid spherical resonator geometry is illustrated in the inset, where the inner Si radius is $r_{Si}$=0.43μm, $r_{PbTe}$ = 0.62μm and the outer Si layer radius is $r_{Si}$ = 1.2μm (b) Transmission amplitude and phase spectra at T=143K and T=643K, respectively. Negligible variations in both amplitude and phase are observed, demonstrating that in addition to resonant frequency, both the phase and the amplitude remain the same between these the two extreme temperatures (143K vs. 643K).

Evidently, temperature independent response is manifested by a small FOM of all resonances ($|S_{max}|$ <0.0035) for this spherical resonator. Complete cancelation of the effective thermo-optic effect in the resonator ($dn_{eff}/dT$=0) is limited to ~ ΔT=150K. For example, the MQ mode (green dots, lower panel of Figure 4), was optimized to exhibit zero resonance wavelength shift $d\lambda/dT$=0 for 193K< T < 343K. For comparison, with no thermo-optic correction, a similar sized silicon resonator would have exhibited a Δλ=75nm resonance wavelength shift for the same ΔT=150K. Furthermore, analysis of the temperature dependent electric and magnetic field distributions for all the resonances, reveal that the resonant mode field profiles also remain fixed across the full 500K temperature gradients (see supporting information section 2.1.).

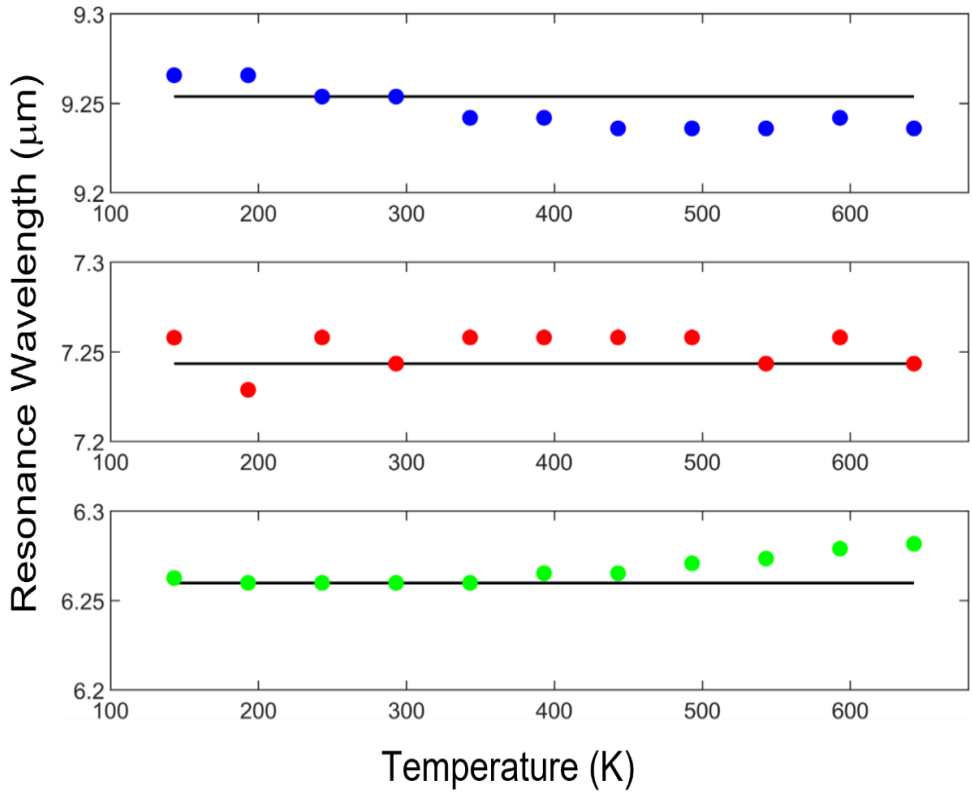

*Figure 4:* Extracted resonance wavelength vs. temperature for MD (blue), ED (red) and MQ (green) resonances in spherical resonators shown in figure 3a. Temperature independent response is manifested by a small FOM of all resonances ($S_{max}$ <0.0035)

## 2. Disk and Cubic Hybrid Mie resonators

Disk and cubic resonators are the most common unit-cells in meta-optics and possess additional degrees of freedom compared to spheres, as their geometry breaks the spherical symmetry. The FDTD scattering spectra, along with the amplitude and phase response of cubic and disk hybrid Mie-resonators are shown in Figure 5. The breaking of spherical symmetry, especially along the z axis compared to the x-y plane, is known to change the dispersion of different Mie resonant modes and is often used to design unidirectional Huygens metasurfaces [39,40]. Here, the different polarization of Mie-resonances leads to different temperature dispersion of resonances which makes it more challenging to simultaneously design temperature invariant response for two or more resonant modes. We focus on designing temperature invariant response around one resonance mode per structure, where canceling the TO dispersion in each structure is optimized for one dipolar mode. The spectra presented in Figures 5a and 5c, demonstrate minor wavelength variations over large temperature span (ΔT=500K), for a particular dipolar mode, of the individual resonator. $S_{max}$ for the fundamental mode in the cubic structure (circled in figure 5a), receives a value of $|S_{max}|$=0.0012 while $|S_{max}|$= 0.0013 is obtained for the 2$^{nd}$ mode in the disk geometry (highlighted by a circle in figure 5c). The field profiles for these structures were also calculated and can be found in the supporting information section 2.2. Furthermore, both structures show minor variations in phase and amplitude (Figures 5b and 5d), between the two extreme temperatures (T=143K and T=643K). These variations are mostly observed around the abrupt changes at the resonance wavelengths and are slightly larger for the disk geometry.

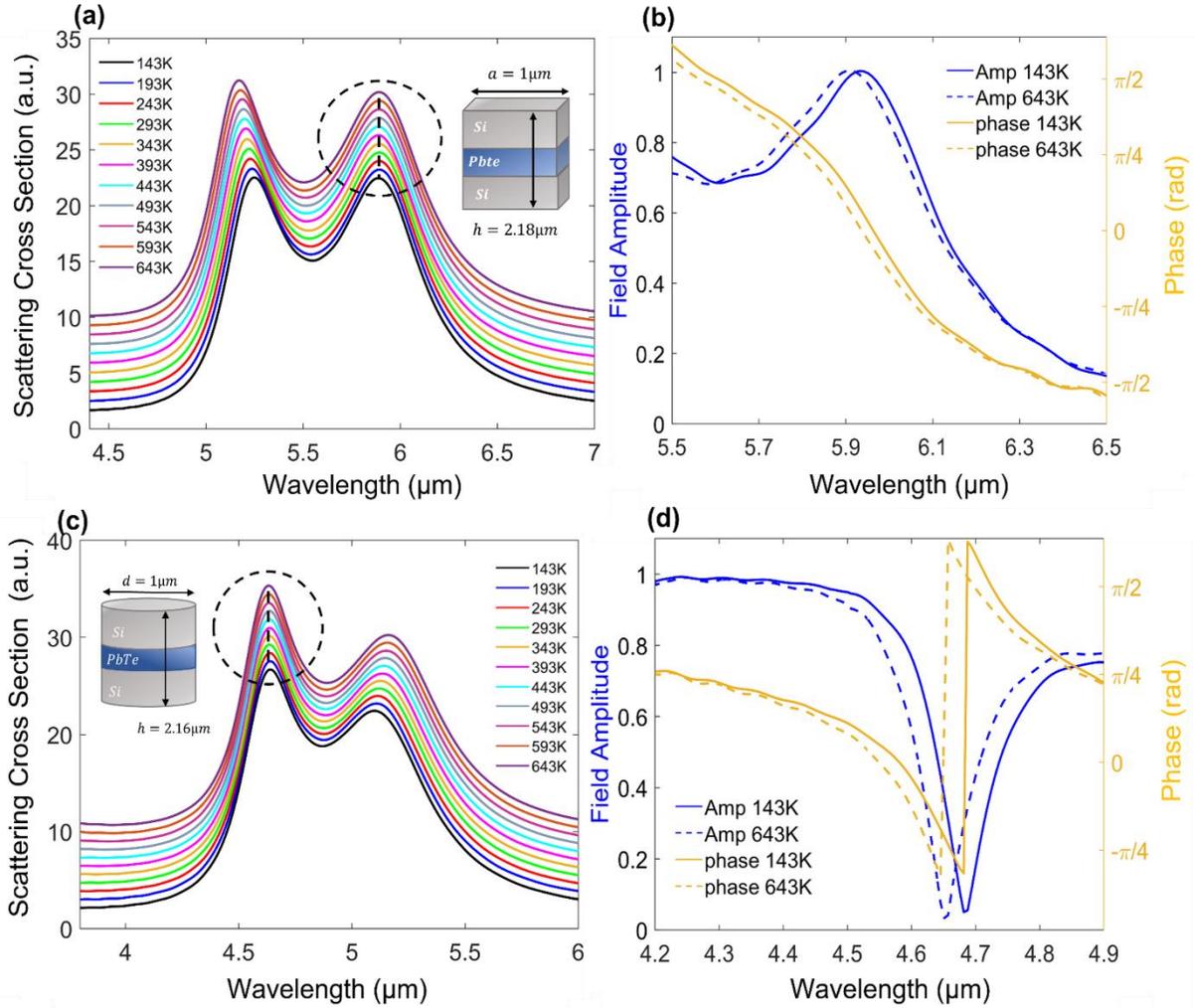

**Figure 5:** Disk and cubic multilayer hybrid resonators. **(a)** Scattering spectra of a cubic resonator demonstrating pinned resonance wavelength behavior across the 143K-643K temperature range. The spectra are vertically shifted along the y-axis for visibility. The hybrid cubic resonator dimensions are $d_{Si}$ = 0.92µm, $d_{PbTe}$ = 0.34µm and $d_{Si}$ = 0.92µm, while the side is a=1µm. The spectra demonstrate temperature independent behavior for the first mode (dashed circle) **(b)** Reflection Amplitude and phase for T=143K (blue) and T=643K (orange), exhibiting minor variations in both the phase and amplitude for the two extreme temperatures. **(c)** Hybrid disk scattering spectra, demonstrating temperature independent behavior for the second mode (dashed circle). The disk dimensions are $d_{Si,1}$ = 0.96µm, $d_{PbTe}$ = 0.24µm and $d_{Si,2}$ = 0.24 µm, while the radius is r=500nm **(d)** Transmission amplitude and phase around the second (ED) mode. Minor variations in both amplitude and phase are visible.

Figure 6. summarizes the temperature dependent resonant wavelength properties of cubic and disk hybrid Mie resonators. Temperature independent response is manifested by near zero values of $S_{max}$ for both structures, showcasing the ability to tailor the unit cell parameters in order to mitigate temperature effects per given mode. Similar to the spherical resonator, complete elimination of the effective thermo-optic response can be achieved for a reduced temperature range of ΔT≈ 200K. This can be seen in Figure 6b, where the disk resonator exhibits zero resonance wavelength shifts (dλ/dT=0), between T=443K and T=643K. Similarly, the parameter sweep process can be further optimized for any of light's degrees of freedom (e.g., the phase, amplitude polarization, angular momentum etc.).

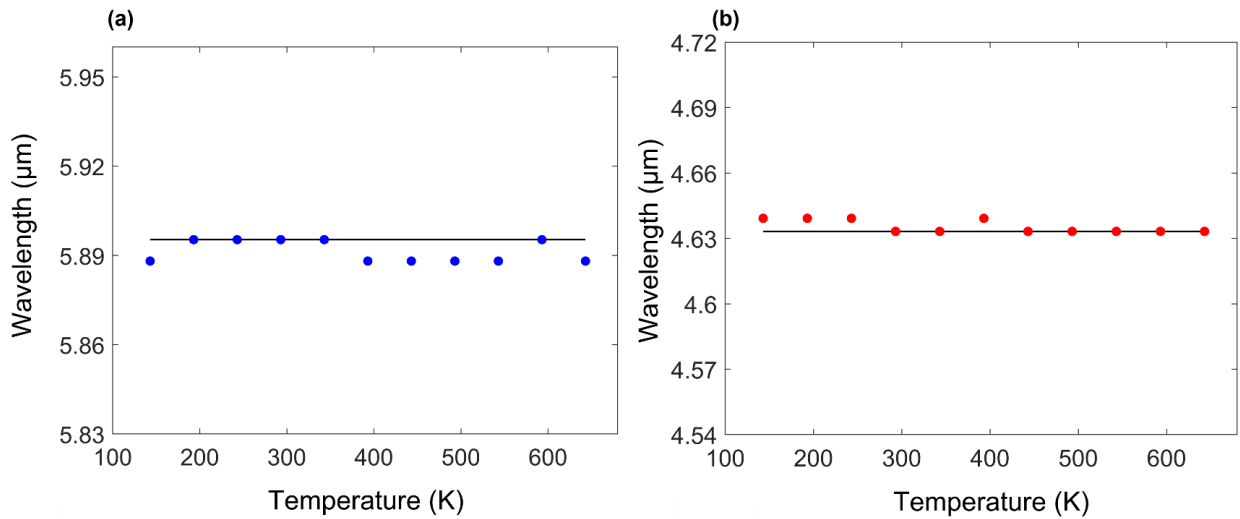

**Figure 6:** Extracted resonance wavelength vs. temperature for **(a)** Fundamental mode in the cubic resonator shown in Figure 5a and **(b)** The 2nd mode of the disk resonator shown in Figure 5C. Temperature independent response is manifested by near zero values of $S_{max}$ ($S_{max}$ ≈ 0.0013)

### 3. Hybrid disk Metasurfaces

Following the study of temperature invariant response in single meta-atom unit cells, we move forward to implementing full metasurface arrays. Such meta-optic components could be integrated into nanophotonic and electro-optic devices (filters, beam shaper, lenses, etc.) providing stable, robust and temperature independent response.

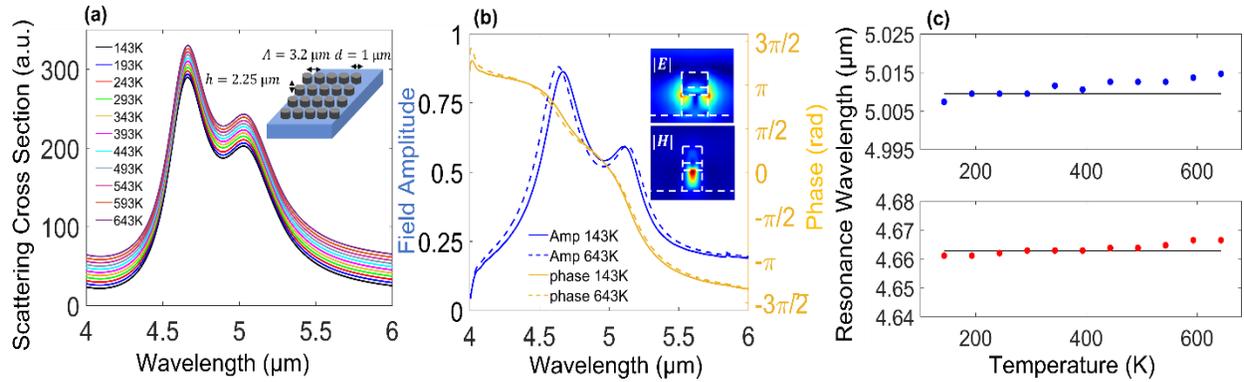

**Figure 7:** (a) Hybrid disk metasurface scattering spectra demonstrating near perfect temperature invariant behavior. The metasurface geometry is illustrated in the inset (b) Hybrid disk metasurface reflection spectrum, demonstrating pinned behavior of the amplitude and phases of T=143K vs T=643K. Field profiles of the fundamental mode are depicted in the inset (c) Extracted resonance wavelength vs. temperature for the Fundamental (blue, upper panel) and the 2nd (red, lower panel) modes of the hybrid disk metasurface. Temperature independent response is manifested by near zero values of ($S_{max}$ ≤0.001)

Figure 7a presents FDTD calculated spectra for a metasurface comprised of hybrid disk resonator unit cells. We selected $BaF_2$ as the substrate material, due to its low thermo-optic response ($dn/dT \approx -19 \cdot 10^{-6}[1/K]$) [41], low refractive index, as well as a wide transparency window in the visible to MIR range [42]. The metasurface layout along with the lattice constant ($\Lambda$=3.2µm) are shown in the inset of Figure 7a. Since the periodicity (lattice constant) of the structure is smaller than the free space wavelength of the incident light, the overall scattering properties are mostly inherited by the single unit cell resonator [39,43]. For very small lattice constants, the inter-particle interactions would be more significant, however these conditions were not considered here.

Our results demonstrate that the scattering cross section, amplitude, resonance wavelength, as well as the phase, for the two fundamental modes – all maintain their RT values (Figures 7a and 7b). Minor resonance wavelength shifts of $\Delta\lambda_{max}$ < 5nm are observed for the two resonance modes, as can be seen in Figure 7c. The corresponding $S_{max}$ values for the two fundamental modes are $S_{max}$ = 0.001 and $S_{max}$ = 7.72 x $10^{-4}$, for 143K < T < 643K, respectively. Perfect temperature invariant performance (Δλ=0 ,$S_{max}$=0) is observed between T=293K and T=393K (lower panel in Figure 7C), while maximum resonance shift as small as Δλ=0.9nm, is obtained for a wider temperature range ΔT=250K (243K<T<493K). As we noted above, a similar silicone metasurface with no TO correction, would have exhibited a Δλ =125nm for the same temperature gradient ΔT=250K.

Figure 7b presents the reflection phase, demonstrating temperature invariant full 2π phase coverage across the two fundamental modes. Field profiles of the fundamental dipolar mode – electric and magnetic - are presented in figure 7b, exhibiting the typical field distribution of MD, similar to the field patterns of single disk resonators (see supporting information section 2.2). Spanning 2π phase coverage is fundamental to meta-optic design as it allows to implement a plethora of optical functionalities. The capability to maintain temperature independent resonant wavelength along with 2π phase coverage in a realistic meta-optic device, showcases the strength of the proposed design and demonstrates the potential to implement TO dispersion engineering in nanophotonics devices.

**Conclusions**

In this work we propose a method to eliminate thermally induced shifts in the optical properties of nanophotonic and meta-optic devices. Our approach is based on compound unit cells composed from at least two materials with opposite TO coefficients allowing to engineer near zero effective TO coefficient ($dn_{eff}/dT≈0$). We demonstrate temperature invariant response for 1D structures such as Bragg mirrors, as well as various 2D single resonator geometries. We demonstrate very small variations in the resonance frequency ($S_{max}$<0.001), amplitude and phase of these resonators, across large temperature gradients spanning 500K degrees. The reduction in thermally induced optical shifts are at least an order of magnitude smaller than the shifts of similar resonators with no TO shift correction. The temperature invariant capabilities of full metasurface arrays, surpass single resonators, due to the reduction in the scattering channels and increase in resonance quality factors. Hybrid disk metasurfaces exhibited near perfect temperature invariant response with the effective thermo-optic response of this structure is at the scale of $d\lambda_{eff}/dT \approx 5$ pm/K, for the temperature range 143K < T < 643K. Peak performance for this metasurface was achieved $S_{max} = 7.72 \times 10^{-4}$, which is also the best value reported here for any of the of the studied nanophotonic structures. It should be noted, that for relaxed temperature gradient conditions of $\Delta T \approx 150K$, we achieved perfect zero effective thermo-optic response with $S_{max} \approx 0$, completely eliminating TO effects.

In summary, our approach for temperature invariant engineering in single antenna resonators and meta-optics devices, provides additional degrees of freedom to optical design capabilities, allowing to compensate for, and optimize TO dispersions effecting resonance wavelength, amplitude and phase. We demonstrate that near perfect temperature invariant performance can be achieved for large temperature variations spanning 500K degrees, which would have otherwise significantly altered the optical properties of the device. These results may pave the way for temperature invariant response in a vast array of applications which are sensitive to temperature variations, enhancing efficiency, stability, and performance.

**Acknowledgements**

We thank the Israel Science Foundation (ISF) for funding this work under grant no. 2110/19

# Supporting Information

# Temperature Invariant Metasurfaces

Shany Zrihan Cohen[1,2], Danveer Singh[1,2], Sukanta Nandi[1,2] and Tomer Lewi[1,2]

[1]Faculty of Engineering, Bar-Ilan University, Ramat-Gan 5290002, Israel
[2]Institute for Nanotechnology and Advanced Materials, Bar-Ilan University Ramat-Gan 5290002, Israel
*Corresponding authors: tomer.lewi@biu.ac.il

## 1. The thermo-optic effect

The thermo-optic (TO) effect describes the variation of the refractive index with temperature, and can be defined in the transparent regime as [1]:

$$(1) \quad 2n\frac{dn}{dT} = (n_\infty^2 - 1)\left(-3\alpha R - \frac{1}{E_{eg}}\frac{dE_g}{dT}R^2\right)$$

where n, $n_\infty$ and T are the refractive index, the high frequency refractive index and temperature respectively, α is the linear thermal expansion coefficient, and $R = \frac{\lambda^2}{\lambda^2 - \lambda_{ig}^2}$ where $\lambda_{ig}$ is the wavelength corresponding to the temperature-invariant isentropic bandgap [1], and $E_{eg}$ is the temperature-dependent excitonic bandgap.

As the relation in Eq 1. suggests - the TO coefficient encapsulates contributions from several physical mechanisms such as thermal expansion, excitonic and phonon excitations. For semiconductors and dielectrics, operating in the normal spectral regime (for λ>$\lambda_{ig}$ and therefore R>0) – the dominant contributor is the excitonic bandgap, typically by two orders of magnitude (α is typically ~$10^{-6}$/°K while $dE_{eg}/dT$ ~$10^{-4}$eV/°K). For vast majority of semiconductors (Silicon included), the excitonic bandgap decreases with temperature ($\frac{dE_{Eg}}{dT} < 0$) and since it is multiplied by another negative term (-1/ $E_{eg}$), it follows that the TO coefficient is usually positive. Indeed, this is the case for the vast majority of semiconductors including silicon. Interestingly, the lead chalcogenide family PbX (X=Te, Se, S) - exhibit an anomalous bandgap energy dispersion with temperature $dE_g/dT$>0 and hence also a similar behavior for the excitonic bandgap dispersion $dE_{eg}/dT$>0,i.e., the bandgap increases when temperature increases [2] and hence the refractive index in lead chalcogenides decreases with temperature, in contrast to most materials. This anomalous negative TO coefficient in PbTe is the enabling component for the TO dispersion engineering presented in this work.

Throughout this work we used TO coefficients as detailed in ref [1], where $dn_{Si}/dT$ = 2.5 [$10^{-4}/{}^{0}K$], $dn_{PbTe}/dT$ = -13.5 [$10^{-4}/{}^{0}K$]. The chromatic dispersion of Si at room temperature (RT) was taken from Palik [3] and the dispersion relation of PbTe is given by [4,5] :

$$(2)\ n_{PbTe}(\lambda) = \sqrt{1 + \frac{30.586\lambda^2}{\lambda^2 - 2.0494} - 0.0034832\lambda^2}$$

When calculating thermal and TO effects, we assumed materials and components have reached equilibrium and did not consider any transient effects. We also did not consider any nonlinear effects.

2. Field Profiles

2.1 Multilayer core-shell Spherical Resonators

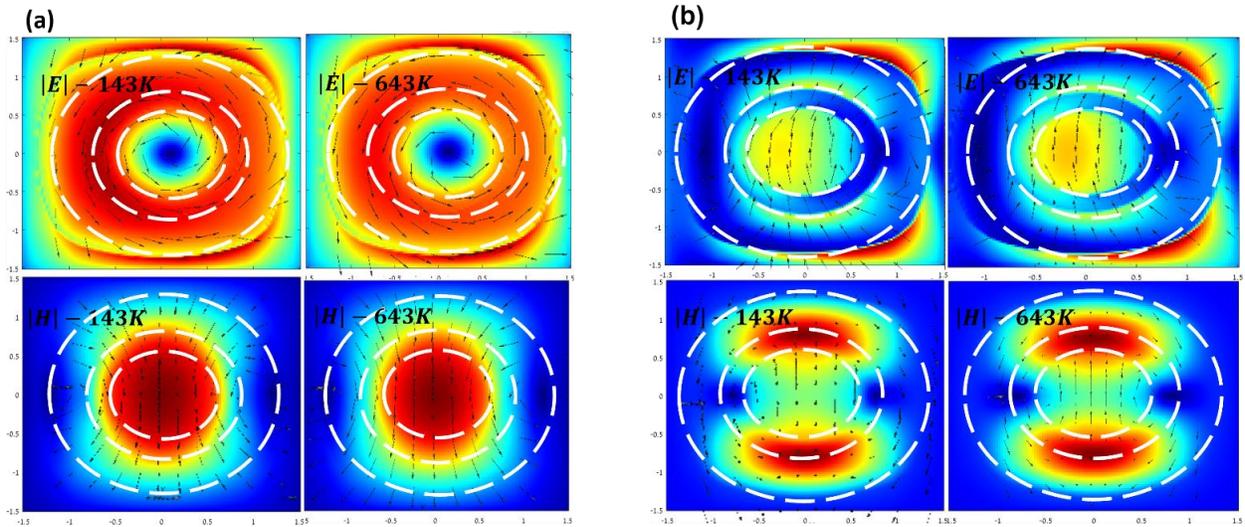

Figure S1. Electric and magnetic field profiles along the xz plane for the first two fundamental resonances, calculated at the lowest (143K) and highest (643K) temperatures (a) Fundamental MD resonance at $\lambda \cong 9.25\mu m$ (b) ED Mie resonance at $\lambda \cong 7.22\mu m$. The 3-layer hybrid sphere dimensions are $r_{Si}$ =0.43μm, $r_{PbTe}$ = 0.62μm and the outer Si radius is $r_{Si}$ = 1.2μm.

The field distributions of the first two resonant modes of the multilayer hybrid spherical resonators, presented in the main text are shown in Fig S1. The first resonance at $\lambda \cong 9.25\mu m$ (Fig S1a) shows a clear distribution of a magnetic dipole (MD) mode. The second resonance at $\lambda \cong 7.22\mu m$ (Fig S1b) resembles an electric dipole (ED), however the field pattern is more complex compared to a single sphere, due to the additional shell layers. Interestingly, field profiles of both modes seem almost identical in the two extreme temperature values $143K \: and \: 643K$.

## 2.2 Multilayer Cubic and Disk Hybrid Resonators

Figure S2 and S3 present the electric and magnetic field patterns of the cubic and disk structures at the fundamental MD resonance, at the lowest (143K) and highest (643K) temperatures.

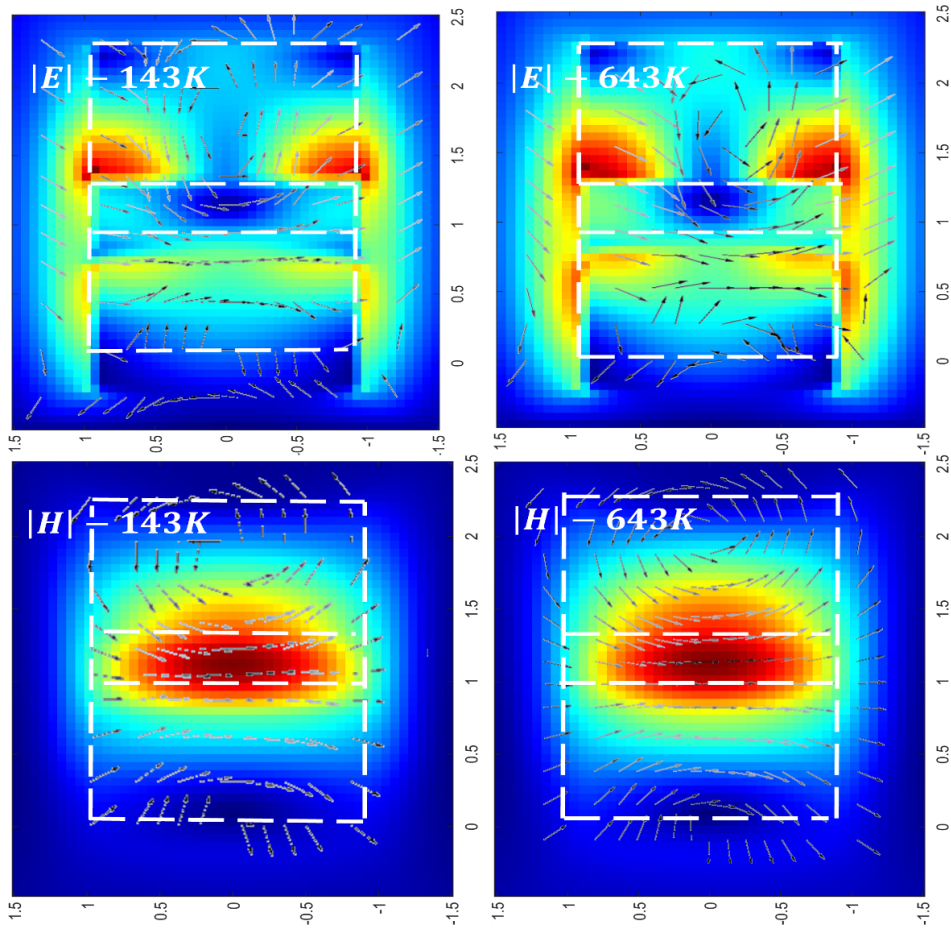

Figure S2 Electric and magnetic field patterns (xz plane) of the cubic resonator at the first resonance λ=5.86μm. the cubic geometry is: h$_{Si}$ = 0.92μm, h$_{PbTe}$ = 0.34μm, length and width at the x-y plane are 1μm each

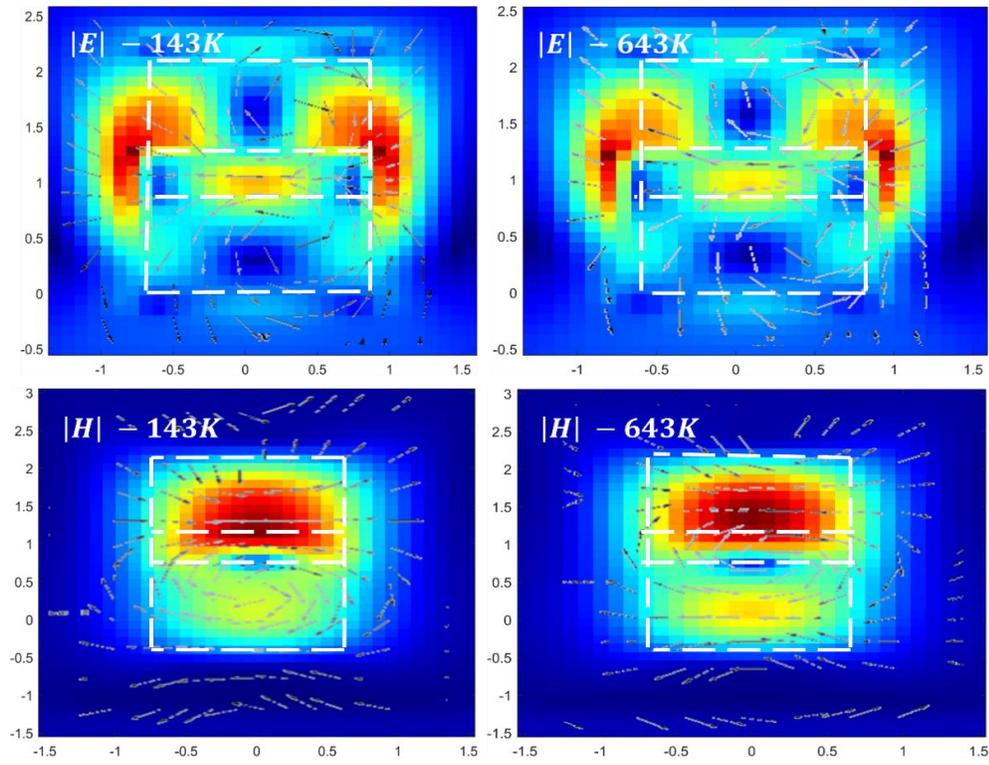

Figure S3  Electric and magnetic field patterns (xz plane) of the disk resonator at the second resonance λ=4.71µm. The disk geometry is: $h_{Si}$ = 0.96µm, $h_{PbTe}$ = 0.24µm, r=0.51µm.

## 2.3 Metasurface comprised of Disk Hybrid Resonators

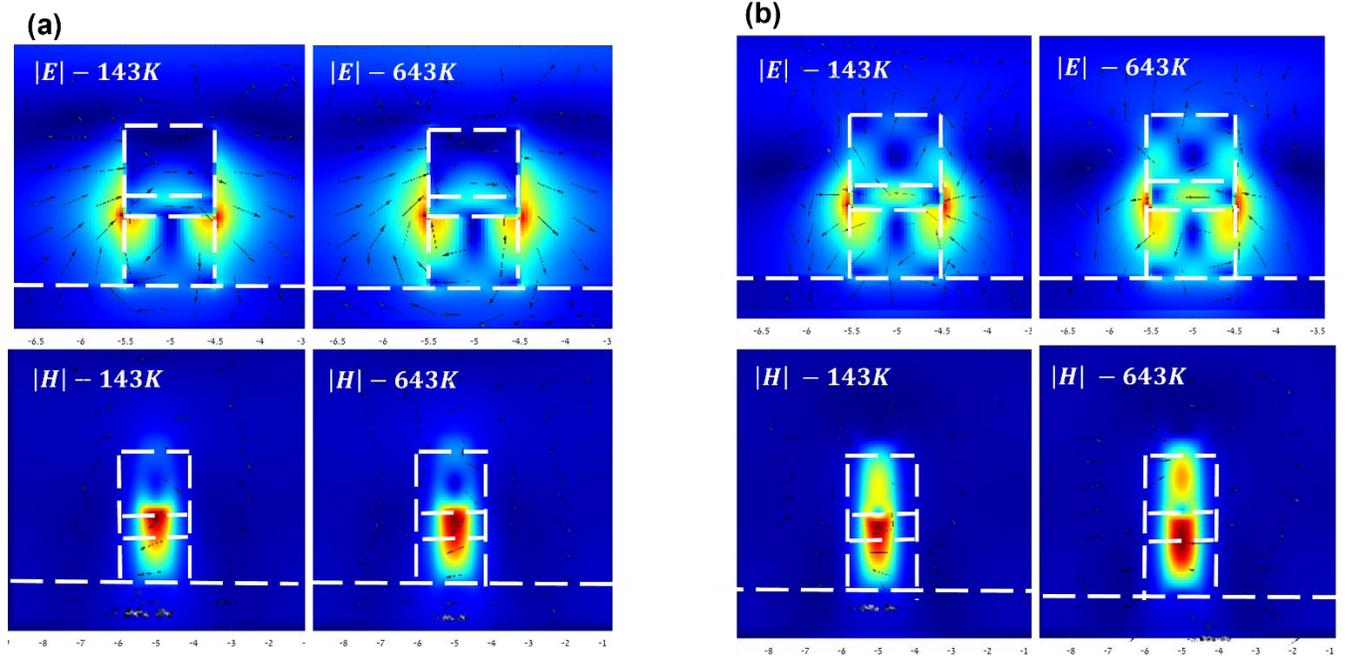

Figure S4: Electric and magnetic field patterns for the first two resonances (in the XZ plane?) of the hybrid unit cell disk metasurface. The images compare the filed profiles at the lowest and highest temperatures – (a) Fundamental resonance at λ≈5µm demonstrating MD pattern (b) second resonance at λ≈4.66µm demonstrating ED pattern. The hybrid disk resonator geometry is $h_{Si}$ =1µm, $h_{PbTe}$ = 0.25µm, r = 1µm and the periodicity is Λ = 3.2µm

# 3. Lumerical Finite Difference Time Difference Simulation Settings

Finite difference time domain calculations were performed using the Lumerical Solutions FDTD solver, Version 8.25.2647. New materials with the required dispersion data were imported to Lumerical data set. The source is a plane wave

**Simulations of Cross Section Scattering**

These simulations were computed using the 'cross-section' analysis object. The simulation region consisted of each of the structures surrounded by air. A non-uniformal conformal mesh was used. A mesh size of at least 10x smaller than the minimum wavelength in the material was used with boundary conditions of perfectly matched layer.

**Simulations of Reflection and Transmission of a full Metasurface**

These simulations were computed using the 'Grating S parameters' analysis object. The simulation region consisted of a full metasurface comprised of each structure surrounded by air.

PML boundary conditions were used in the radiation axis $\hat{z}$, while periodic symmetric and antisymmetric boundary conditions were used in the $\hat{x}, \hat{y}$ directions, respectively. The boundary conditions are selected with respect to the polarization direction and dictates the electric and magnetic field components which are zero at the plane of symmetry [6–8]